# Systematic analysis of the pp collisions at LHC energies with Tsallis function

Murad Badshah[1], Muhammad Waqas[2,†], Ahmed M. Khubrani[3], Muhammad Ajaz[1,*]

[1] Department of Physics, Abdul Wali Khan University Mardan, Mardan 23200, Pakistan
[2] School of Nuclear Science and Technology, University of Chinese Academy of Sciences, Beijing 100049, China,
[3] Department of Physics, Faculty of Science, Jazan University, 45142 Jazan, Saudi Arabia

†Corresponding author: waqas_phy313@ucas.ac.ch (M. Waqas)

*Corresponding author: ajaz@awkum.edu.pk (M. Ajaz)

## Abstract

This work focuses on the study of identified ($\pi^\pm$, $k^\pm$, p, and $\bar{p}$), strange hadrons ($k_s^0$, $\Lambda$, $\bar{\Lambda}$, $\Xi^+$, $\Xi^-$), recorded by CMS, and light nuclei and their anti-nuclei (d, $\bar{d}$, t, $\bar{t}$, $^3He$ and $^3\overline{He}$), recorded by ALICE, at $\sqrt{s} = 0.9$ TeV, 2.76 TeV, 7 TeV and 13 TeV in pp collision at mid rapidities. The $p_T$ distributions of these particles are analyzed using the Tsallis model, which fits the experimental data very well. Several important parameters for studying the characteristics of the medium produced during such collisions are extracted. The effective temperature (T) increases monotonically with increasing particle mass and also with increasing collision energy. The non-extensivity parameter (q) decreases with the mass of the particle. For heavier particles, greater T and smaller q mean that they decouple early from the system and attain equilibrium quickly compared to lighter ones. Furthermore, with an increase in collision energy, the multiplicity parameter $N_0$ increases.

## Introduction

The collision of relativistic nuclei produces hot and dense matter, similar to the conditions prevailing in the early universe, called Quark Gluon Plasma (QGP) [1-5], where quarks and gluons have asymptotic freedom. One of the primary aims of experimental particle physicists at the Large Hadron Collider (LHC) and Relativistic Heavy Ion Collider (RHIC) is to create this state of matter and to investigate its behaviour and different kinds of properties which include space-time evolution of the medium, variation of thermal effects, flow effects, size of the medium with varying collision energy [6-8], the centrality of the colliding nuclei [7], rapidity [9,10] and size of the colliding nuclei [11]. During its evolution, the QGP comes across two main stages, the chemical freezeout stage and the kinetic or thermal freezeout stage, and their corresponding temperatures are called chemical freezeout temperature and kinetic freezeout temperature, respectively.

The European Organization for nuclear research (CERN) announced the discovery of QGP in 2000 in a nucleus-nucleus collision, this was an indirect discovery based on many signatures, including strangeness enhancement, charmonium suppression, or charmonium melting, jet quenching, direct photons, and elliptic flow, etc. We can use the transverse momentum of the particles to extract the lost information, such as the transverse excitation degree and dynamic expansion of the collision system, due to the short lifetime of the QGP [12].

Temperature is one of the main factors which plays a vital role in forming QGP. In the system evolution, there are different kinds of temperatures, namely the chemical freezeout temperature, effective temperature, and kinetic freezeout temperature, and they appear at different stages during the evolution system and correspond to different characteristics. In the present work, we only deal with the effective temperature, which appears just before decoupling the particle from the system. Effective temperature includes the system's thermal as well as flow effects and can be given as $T = T_0 + m_0 <\beta_T>^2$ [13], where T is the effective temperature, $T_0$ is the kinetic freezeout temperature and $<\beta_T>$ is the transverse flow velocity. The scenario of the particle's decoupling (effective temperature and kinetic freezeout temperature) undergoes the single, double, triple, and multiple decoupling scenarios in different studies, which is very knotty and need to be decoded. As a matter of fact, the freezeout of the particle has a complicated structure according to various studies because the excitation function of the system gives multiplex results in different studies [14-18]. In the present study, we have analysed the $p_T$ distributions of different particle species at different energies to dig out the accurate freezing-out of the particles and to study the subordination of the effective temperature (T) and other relevant parameters on the collision energy.

As stated above, QGP has been discovered in a nucleus-nucleus collision, and many signatures relevant to such a state of matter have also been observed in small systems like p-p collisions. Some of the observed signatures in high multiplicity p-p events include strangeness enhancement [19], production of a large number of particles [20], equivalent effective temperature, kinetic freezeout temperature, and flow velocity to those obtained in nucleus-nucleus collisions [21, 22] and multiparticle ridge-like correlations [23]. p-p collisions are the baseline for studying AA or p-A collisions.

The particles' transverse momentum ($p_T$) spectra contain some crucial information about the thermal excitation of the system and its expansion. The high $p_T$ particles are assumed to have crucial information about jet quenching, a phenomenon in which reduction occurs in a jet's energy due to its interaction with the medium. In QGP [24], the low $p_T$ particles produced due to multiple interactions in the medium and following the exponential distribution functions provide the thermal information of QGP [25]. In contrast, the intermediate $p_T$ particles provide information about quark recombination or coalescence in the collision process [26]. This implies that the $p_T$ distribution of the produced particles or hadrons encodes handy information about the collision process and the medium just created after the collision. The detectors fixed at colliders are not so powerful and highly advanced yet to measure all the quantities relevant to the QGP, including the system's temperature, size or volume of the system and flow velocity etc. Therefore, various statistical models can be incorporated to fit with experimental data of $p_T$ distribution of the observed hadrons and to extract the relevant parameters. In this work, the

p_T distribution of identified particles, strange particles, light nuclei and their anti-nuclei have been described using the Tsallis model [27-31], and the effective temperature, nonextensive parameter and kinetic freezeout volume of the system have been derived.

The rest of the paper is organized as follows: Section 2 describes the experimental data and model used in this research. Section 3, describes our results, while section 4, presents the conclusion of the present work.

## 2. Experimental data and model

The experimental data describing the p$_T$ distribution of identified particles at $\sqrt{s} = 0.9$ TeV, 2.76 TeV, 7 TeV and 13 TeV are taken from [32] and [33], for the strange particles the experimental data at $\sqrt{s} = 0.9$ TeV and 7 TeV are taken from [34] while for light nuclei and their anti-nuclei at $\sqrt{s} = 0.9$ TeV. 2.76 TeV and 7 TeV the data are taken from [35].

The most widely studied observables in high-energy particle collisions are the transverse momentum spectra of the produced particle. To describe the p$_T$ spectra, the exponential function, called the Boltzmann-Gibbs exponential function given in Eq. 1, has been used for decades.

$$f(E) \sim \exp\left[\frac{-E}{T}\right] \qquad (1)$$

Where E is the energy and T is the corresponding temperature. However, the Boltzmann-Gibbs exponential function fails to describe the high p$_T$ regimes, i.e., at p$_T$ > 3 GeV/c. This is because different physical processes govern particle production in this regime, and particle production is assumed to be governed by non-thermal and non-equilibrium perturbative QCD processes. Instead of the exponential distribution, a power law distribution is found to be more appropriate in such a regime [36].

In order to formulate the power law distribution, to cover the whole regime, an improvement has been made in the statistical picture, in which the Boltzmann-Gibbs theory is generalized in the context of non-extensive statistics, and what has been obtained in this generalization is called non-extensive Tsallis distribution function. The Tsallis function used to describe p$_T$ spectra of the particles in its simplest form is given in Eq. 2 or Eq. 3.

$$f(p_T, m_T, T, q) = 2\pi C\, p_T \left[1 + (q-1)\frac{m_T}{T}\right]^{-1/(q-1)} \qquad (2)$$

$$\frac{d^2N}{N_{ev}\, dp_T\, dy} = 2\pi C\, p_T \left[1 + (q-1)\frac{m_T}{T}\right]^{-1/(q-1)} \qquad (3)$$

Where T is the effective temperature, q is called the non-extensivity parameter, and if it tends to unity, the system's thermalization level increases. At q = 1, Eq. 2 or Eq. 3, the Tsallis function approaches to Boltzmann-Gibbs exponential function, C is the normalization constant and is given as $C = gV/(2\pi)^3$. Where g represents the degeneracy factor, and V is the kinetic freezeout volume of the particles. The $m_T$ is the transverse mass given by $m_T = \sqrt{p_T^2 + m_0^2}$, where $m_0$ is the rest mass of the particle.

There are many forms of Tsallis function (model) that have been used in the literature, the one used in our coding, called thermodynamically consistent Tsallis function [37], is given as,

$$\frac{d^2N}{N_{ev}\, dp_T\, dy} = 2\pi C\, p_T\, m_T \left[1 + (q-1)\frac{m_T}{T}\right]^{-q/(q-1)} \qquad (4)$$

## 3. Results and Discussion

Fig. 1 is obtained by fitting the Tsallis model with experimental data for $\sqrt{s} = 0.9\ TeV$, $2.76\ TeV, 7\ TeV\ and\ 13\ TeV$ in pp collision, the solid lines represent the model fitting and the symbols are used for experimental data points. Different symbols are used for the representation of different particles. The $p_T$ spectra of identified particles ($\pi^\pm$, $k^\pm$, p and $\bar{p}$), strange hadrons ($k_s^0$, $\Lambda$, $\Xi^-$) and light nuclei and their anti-nuclei (d, $\bar{d}$, t, $\bar{t}$, $^3He$ and $^3\overline{He}$) at 0.9, 2.76, 7 and 13 TeV are shown in Panel (a), (b), (c) and (d) respectively. Panel (a) represents the $p_T$ distribution of $\pi^\pm$, $k^\pm$, p, $\bar{p}$, $k_s^0$, $\Lambda$, $\Xi^-$ d, while panel (b) represents the $p_T$ distribution of $\pi^\pm$, $k^\pm$, p, $\bar{p}$, d, and $\bar{d}$. Panel (c) and (d) represent the $p_T$ spectra of $\pi^\pm$, $k^\pm$, p, $\bar{p}$, $k_s^0$, $\Lambda$, $\Xi^-$, d, $\bar{d}$, $^3He$, $^3\overline{He}$, t, $\bar{t}$, and $\pi^\pm$, $k^\pm$, p, $\bar{p}$ respectively. The plots show that the Tsallis model agrees well with experimental data. To enhance distribution visibility, some $p_T$ spectra are scaled with differnt scaling factors, written in the third column of Table 1. The bottom panel of each plot is used to show data by the fit ratio (Data/Fit), which measures the deviation of the fit line (model) from experimental data. The related parameters and the values of $\chi^2$/NDF are tabulated in table 1.

Fig. 2 shows the behavior of effective temperature (T) with the mass of the particle and center of mass energy ($\sqrt{s}$). Different symbols represent T at different energies, while the trend of T from left to right shows its dependence on the particle's rest mass. It is reported that T is mass-dependent. The heavier the particle is, the larger is the effective temperature. This result validates [38-41]. This dependence becomes more pronounced for the hefty particles such as $^3\overline{He}$. Moreovere, T increases with increasing the collision energy, it is because at higher collision energies greater amount of energy is exchanged between the colliding systems which results in higher T. We can see that the dependence of T on collision energy is strong for heavier particles but comparatively weak for light particles. For instance, the values of T for lighter particles are close to one another at different collision energies, while as the particles get heavier, there are magnificent differences in the values of T at different energies. We want to highlight that T for K$^+$ is larger than K$^-$, which contradicts the assumption of simultaneous freezeout of both kaon species. We believe that the coalescence effect for the constituents of K$^+$ is greater than the constituents of K$^-$ which results in the quick formation and decoupling of K$^+$ than K$^-$ and therefore K$^+$ has greater T than K$^-$. The same result of separate freezeout of particles and their antiparticles is observed in our recent work [18, 42], where we analyzed the light nuclei and their antiparticles, and they freezeout separately due to the coalescence effect.

Though the main theme of the present work and [18, 42] is different, there is still a non-simultaneous freezeout of the particles and antiparticles.

*Table 1. The values of different free parameters, normalization constant and $\chi^2$ for identified particles, strange hadrons, light nuclei and their anti-nuclei at different collision energies obtained from the Tsallis model are tabulated here.*

| Energy [TeV] | Particle | Scaling factor | T [MeV] | q | $N_0$ | $\chi^2$/ NDF |
|---|---|---|---|---|---|---|
| 0.9 | $\pi^+$ | 0.01 | 55.68±1.31 | 1.189±0.002 | 31.96±1.00 | 37.3780/20 |
| | $\pi^-$ | 0.002 | 55.68±1.00 | 1.191±0.001 | 31.36±1.00 | 38.7076/20 |
| | $K^+$ | 05 | 81.13±0.80 | 1.151±0.002 | 3.75±0.10 | 2.8745/15 |
| | $K^-$ | 01 | 70.13±0.80 | 1.167±0.002 | 3.76±0.08 | 3.6842/15 |
| | $K_s^0$ | 700 | 83.95±1.00 | 1.139±0.001 | 12.89±0.60 | 20.1932/22 |
| | p | 120 | 89.24±0.7 | 1.112±0.001 | 1.71±0.04 | 14.3054/25 |
| | $\bar{p}$ | 40 | 89.24±0.7 | 1.112±0.001 | 1.61±0.04 | 33.1687/25 |
| | $\Lambda$ | 8×10³ | 104.02±1.00 | 1.097±0.001 | 6.185±0.2 | 8.9475/22 |
| | $\Xi^-$ | 5×10⁵ | 106.31±0.6 | 1.096±0.002 | 0.685±0.1 | 6.9300/20 |
| | d | 05 | 120.01±0.9 | 1.058±0.002 | 0.0029±0.0002 | 0.0559/01 |
| | $\bar{d}$ | 01 | 120.01±0.9 | 1.058±0.002 | 0.0029±0.0002 | 1.2516/01 |
| 2.76 | $\pi^+$ | 02 | 63.12±1.00 | 1.184±0.002 | 38.54±1.00 | 6.8610/20 |
| | $\pi^-$ | 01 | 63.12±1.00 | 1.184±0.002 | 38.01±1.00 | 7.6321/20 |
| | $K^+$ | 100 | 90.13±1.00 | 1.151±0.001 | 4.84±0.05 | 5.5083/15 |
| | $K^-$ | 50 | 80.13±1.00 | 1.169±0.001 | 4.86±0.05 | 9.4057/15 |
| | p | 1.5×10³ | 96.44±1.00 | 1.124±0.001 | 2.16±0.03 | 35.8289/25 |
| | $\bar{p}$ | 700 | 96.44±1.00 | 1.122±0.001 | 2.11±0.03 | 30.4199/25 |
| | d | 8×10³ | 135.01±1.00 | 1.046±0.001 | 0.004±0.0005 | 8.3616/05 |
| | $\bar{d}$ | 4×10³ | 135.01±1.00 | 1.046±0.001 | 0.0038±0.0005 | 6.5521/05 |
| 7 | $\pi^+$ | 0.01 | 72.22±0.70 | 1.171±0.001 | 46.95±1.00 | 56.1243/20 |
| | $\pi^-$ | 1.7×10⁻³ | 72.22±0.70 | 1.171±0.001 | 45.83±1.00 | 44.1055/20 |
| | $K^+$ | 07 | 106.13±0.80 | 1.151±0.001 | 6.13±0.06 | 4.9004/15 |
| | $K^-$ | 01 | 85.13±0.80 | 1.170±0.001 | 6.08±0.06 | 7.7255/15 |
| | $K_s^0$ | 1×10³ | 109.15±1.30 | 1.151±0.001 | 20.9±0.90 | 2.7064/22 |
| | p | 400 | 118.74±1.00 | 1.123±0.002 | 2.71±0.08 | 22.5485/25 |
| | $\bar{p}$ | 100 | 118.74±1.00 | 1.123±0.002 | 2.71±0.08 | 27.8248/25 |
| | $\Lambda$ | 1×10⁴ | 135.24±1.00 | 1.112±0.001 | 11.201±0.01 | 4.1889/22 |
| | $\Xi^-$ | 5×10⁵ | 145.34±1.10 | 1.111±0.001 | 1.34±0.07 | 7.5643/20 |
| | d | 05 | 155.01±1.20 | 1.054±0.002 | 0.0047±0.0002 | 10.5955/19 |
| | $\bar{d}$ | 01 | 155.01±1.20 | 1.054±0.002 | 0.0044±0.0002 | 6.8046/18 |
| | $^3He$ | 40 | 250.43±2.00 | 1.004±0.001 | 0.000004±0.000001 | 0.3471/- |
|   | $^3\overline{He}$ | 08 | 250.43±2.00 | 1.049±0.001 | 0.0000019±0.0000001 | 0.4643/1 |
|   | t | 01 | 250.71±2.00 | 1.003±0.001 | 0.0000028±0.0000001 | 1.1596e-04/- |
|   | $\bar{t}$ | 0.1 | 250.71±0.02 | 1.003±0.001 | 0.0000035±0.0000001 | 4.3612e-04/- |
| 13 TeV | $\pi^+$ | 02 | 80.02±0.90 | 1.155±0.002 | 41.85±0.90 | 61.9276/20 |
|   | $\pi^-$ | 01 | 80.02±0.90 | 1.155±0.002 | 41.65±0.90 | 32.3889/20 |
|   | $K^+$ | 100 | 116.13±0.92 | 1.141±0.001 | 5.55±0.10 | 8.6885/15 |
|   | $K^-$ | 50 | 90.13±0.92 | 1.151±0.001 | 5.27±0.10 | 13.9193/15 |
|   | p | 1×10³ | 130.74±1.00 | 1.111±0.001 | 2.45±0.04 | 17.7516/24 |
|   | $\bar{p}$ | 500 | 130.74±1.00 | 1.111±0.001 | 2.39±0.04 | 14.0792/24 |

*Table 2. The produced particles as a result of p-p collision and their masses in the units of MeV are listed here. The particles are tabulated on the basis of their increasing masses from left to right.*

| Produced particles | $\pi^\pm$ | $K^\pm$ | $K^0_S$ | p | $\Lambda$ | $\Xi^-$ | d | $^3He$ | t |
|---|---|---|---|---|---|---|---|---|---|
| Masses [MeV] | 139.57 | 493.68 | 497.6 | 938.27 | 1115.68 | 1321.71 | 1875.61 | 2808.39 | 2808.92 |

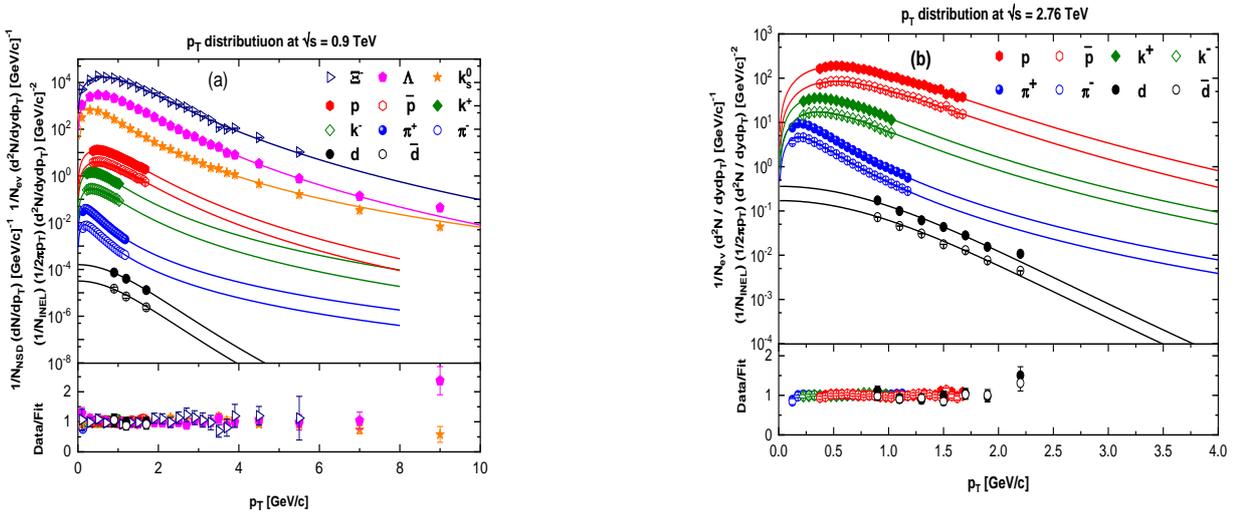

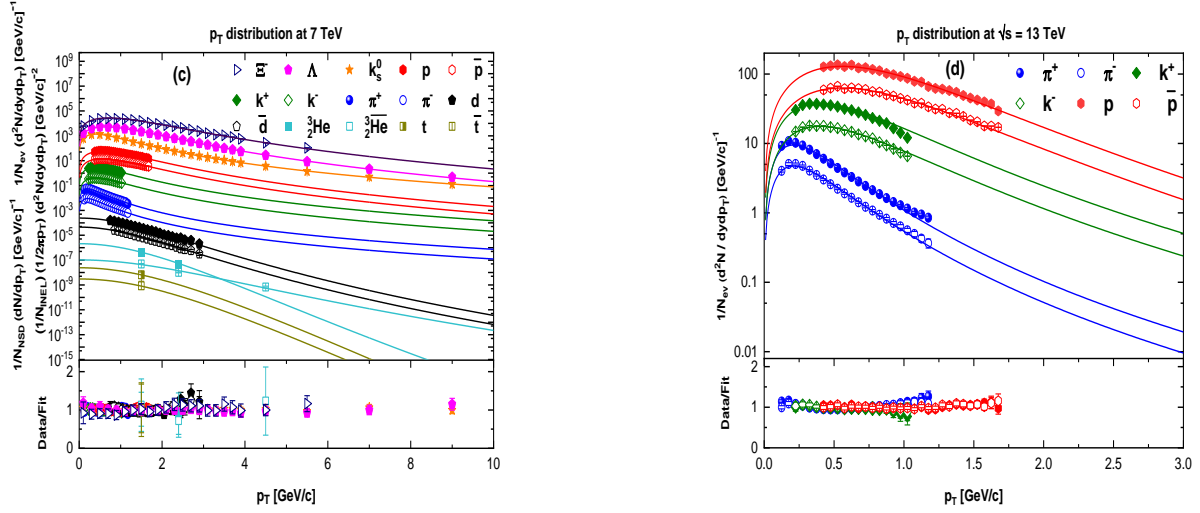

***Fig. 1*** *shows the fit curve for the $p_T$ distribution of identified particles, strange particles, light nuclei, and antinuclei at the different centers of mass energies in pp collision. The symbols show data points, while solid lines represent the fits by the Tsallis model. The lower panel associated with each plot shows the corresponding data by fit ratio.*

Fig. 3 is similar to fig. 2, but the dependence of q is on $m_0$ and $\sqrt{s}$ are shown. One can see that q decreases for heavier particles. However, we did not report any dependence of the parameter q on $\sqrt{s}$. In the present work, there is an inverse correlation between T and the non-extensive parameter (q), as seen in figure 4. The heavier particles have larger T and smaller q, which tend to grab the equilibrium state quickly compared to the lighter ones. The multiplicity parameter ($N_0$) dependence on $m_0$ and $\sqrt{s}$ is presented in fig. 5. $N_0$ increases with an increase in collision energy.

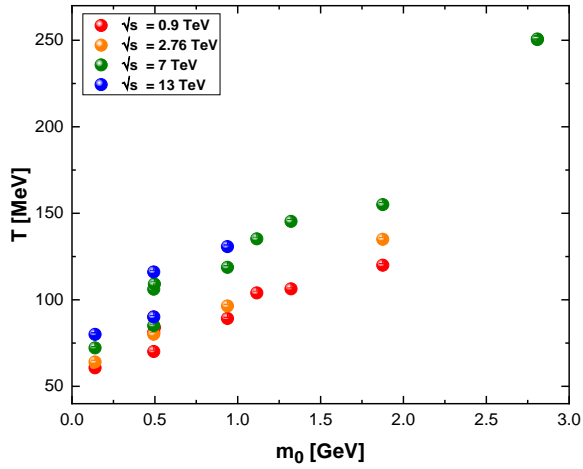

***Fig.2*** *shows the dependence of effective temperature on the masses of the observed particles and collision energy, extracted from the $p_T$ spectra of $\pi^\pm$, $k^\pm$, $p$, $\bar{p}$, $k_s^0$, $\Lambda$, $\bar{\Lambda}$, $\Xi^+$, $\Xi^-$, $d$, $\bar{d}$, $t$, $\bar{t}$, $^3He$ and $^3\overline{He}$.*

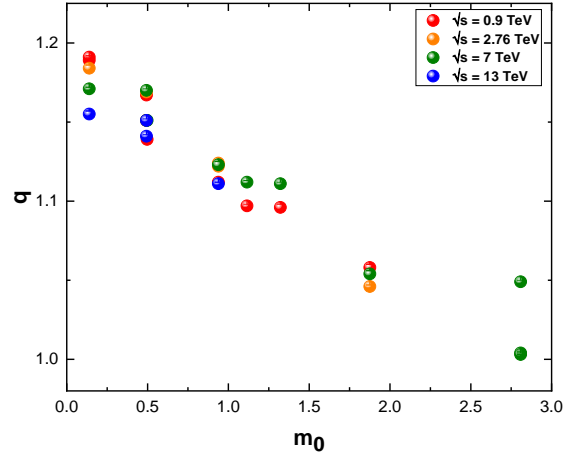

*Fig.3* shows the relationship between masses of the observed particles and non-extensivity parameters at different collision energies, obtained from the $p_T$ distribtions of $\pi^{\pm}$, $k^{\pm}$, $p$, $\bar{p}$, $k_s^0$, $\Lambda$, $\bar{\Lambda}$, $\Xi^{+}$, $\Xi^{-}$, $d$, $\bar{d}$, $t$, $\bar{t}$, $^3He$ and $^3\overline{He}$.

Fig. 4 shows the relation between T and q. The former correlates oppositely with the latter as larger T corresponds to smaller q for the heavier particles. A larger "q" corresponds to non-equilibrium, while its value closer to 1 corresponds to the equilibrium state. It should be noted that the particles come to equilibrium render that they decouple from the system and are detected by the detector. In the analysis of fig. 3, we can say that particles with large "T" attain an equilibrium state quickly and have smaller "q" so the particles freezeout early.

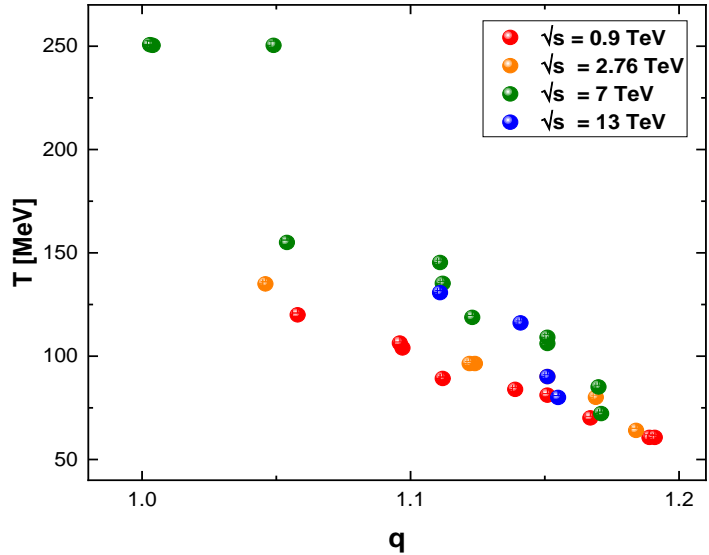

*Fig.4* shows the relationship between effective temperature and non-extensivity parameter at different collision energies, obtained from the $p_T$ spectra of $\pi^{\pm}$, $k^{\pm}$, $p$, $\bar{p}$, $k_s^0$, $\Lambda$, $\bar{\Lambda}$, $\Xi^{+}$, $\Xi^{-}$, $d$, $\bar{d}$, $t$, $\bar{t}$, $^3He$ and $^3\overline{He}$.

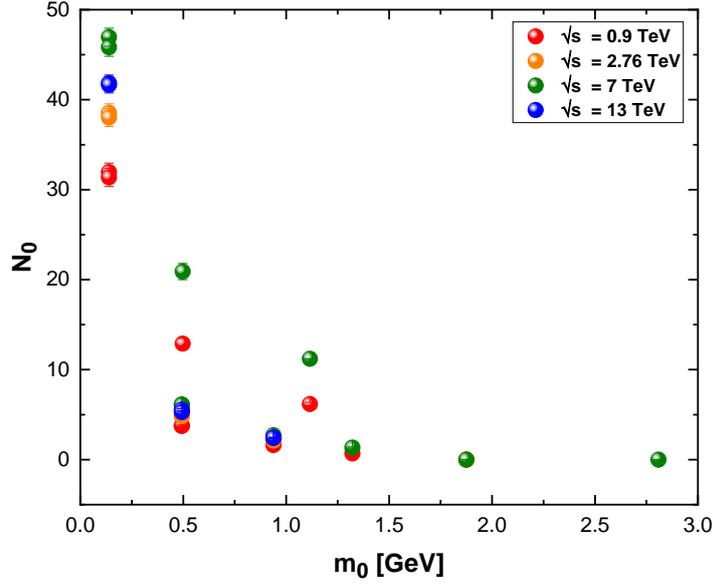

***Fig.5*** *shows the multiplicity parameter ($N_0$) vs. masses of the observed particles at different $\sqrt{s}$, extracted from the $p_T$ distributions of $\pi^\pm$, $k^\pm$, $p$, $\bar{p}$, $k_s^0$, $\Lambda$, $\bar{\Lambda}$, $\Xi^+$, $\Xi^-$, $d$, $\bar{d}$, $t$, $\bar{t}$, $^3He$ and $^3\overline{He}$.*

Fig. 5 displays the multiplicity parameter ($N_0$) dependence on $m_0$ and energy. The symbols show the energy dependence of $N_0$, while the trend of symbols from left to the right shows its dependence on $m_0$. One can see that $N_0$ is larger for the lighter particles and decrease for heavier particles.

## 4.   Conclusion

This research work focuses on the study of identified particles ($\pi^\pm$, $k^\pm$, p, and $\bar{p}$), strange hadrons ($k_s^0$, $\Lambda$, $\bar{\Lambda}$, $\Xi^+$, $\Xi^-$), recorded by CMS, and light nuclei and their anti-nuclei (d, $\bar{d}$, t, $\bar{t}$, $^3He$ and $^3\overline{He}$), recorded by ALICE, at $\sqrt{s} = 0.9$ TeV, 2.76 TeV, 7 TeV and 13 TeV in pp collision at mid rapidities. The Tsallis model has been applied to analyze these particles' $p_T$ distributions. The model fits the experimental data very well. We showed fundamental relations between the different extracted parameters, summarized in the following paragraphs.

We found the effective temperature increases with increasing the energy and the rest mass of the particle, showing the early freezeout of the heavier particles and at higher energies. However, the non-extensivity parameter (q) decreases for heavier particles, which gives the heavier particles a quick approach toward an equilibrium state. Moreover, it has also been found that with increasing collision energy from 0.9 TeV to 7 TeV the multiplicity parameter increases.